\documentclass{DISproc}

\usepackage{slashed}

\begin{document}
%------------------------------------
\title{From Jet Counting to Jet Vetoes}

%for single authors the superscripts are optional
\author{{\slshape Peter Schichtel$^{1,*}$, Christoph Englert$^2$, Erik Gerwick$^3$, Tilman Plehn$^1$, Steffen Schumann$^3$}\\[1ex]
  $^1$ Institut f\"ur Theoretische Physik, Universit\"at Heidelberg, Germany\\
  $^2$ Institute for Particle Physics Phenomenology, Durham University, United Kingdom\\
  $^3$ II. Physikalisches Institut, Universit\"at G\"ottingen, Germany\\
  $^*$ \emph{Speaker} }

% please enter the contribution ID for the DOI
\contribID{xy}

\doi  % if there is an online version we will register DOIs

\maketitle

\begin{abstract}
  The properties of multi-jet events impact many LHC analysis.  The
  exclusive number of jets at hadron colliders can be described in
  terms of two simple patterns: staircase scaling and Poisson
  scaling. In photon plus jets production we can interpolate between
  the two patterns using simple kinematic cuts.  The associated
  theoretical errors are well under control. Understanding such
  exclusive jet multiplicities significantly impacts Higgs searches
  and searches for supersymmetry at the LHC.
\end{abstract}

\section{Introduction}

In LHC searches jets and their properties play an important role for
our understanding of hadron collisions. Jets in association with $W/Z$
bosons as well as pure QCD jets not only help us to understand the
theory, but also pose important backgrounds to new physics
searches. Currently, the Higgs searches are certainly the most
exciting LHC analysis. In the weak boson fusion (WBF) channel these
searches rely on central jet vetoes, where jet radiation between two
hard tagging jets is forbidden~\cite{Rainwater:1999sd}. This idea is
based on the color structure in WBF processes. Nowadays, for example
the $H \rightarrow WW$ searches are divided into \emph{exclusive} $0,
1$ and $2$ jet bins. Whenever new physics scenarios introduce new
heavy colored particles~\cite{Morrissey:2009tf} their search relies on
jets which appear as decay and radiation jets.  The production scale
for such heavy objects is encoded in the effective mass
$m_{\text{eff}} = \slashed{p}_T + \sum_\text{jets} p_{T\text{,jet}}$,
which is essentially proportional to the number of jets.

We propose the \emph{exclusive} number of jets $n_{\text{jets}}$ as
the proper observable to study jets at the LHC. If we control this
observable we can in addition use many multi-jet observables, like
$m_{\text{eff}}$, whose uncertainties are otherwise notorious. There
are, however, some issues in the definition of \emph{exclusive} as
compared to \emph{inclusive} multi-jet observables. To gain higher
precision we usually rely on higher order calculations, which in QCD
predict inclusive observables. This means that once we include parton
densities obeying the DGLAP equation any number of collinear jets is
automatically included. On the other hand, exclusive jet bins are
statistically independent. We use {\sc Sherpa}~\cite{Gleisberg:2008ta}
and its CKKW~\cite{Catani:2001cc} algorithm to generate matched LO
events to study exclusive jet cross-section ratios. In general we
observe two distinct patterns: Poisson and staircase scaling.

\section{Scaling patterns}

\subsection{Poisson scaling}

Poisson processes are well known for example when we rely on the
eikonal approximation~\cite{Peskin:1995ev,Laenen:2010uz}. There, the
matrix element factorizes for example from soft photon emission
\begin{align}
  \label{eq:eikonal}
  \mathcal{M}_{n+1} &= g_s T^3 \epsilon^*_{\mu} \bar{u} (q) \frac {
    q^{\mu} + \mathcal{O} (\slashed{k}) } { q k + \mathcal{O} (k^2) }
  \; \mathcal{M}_n \; .
\end{align}
This relation can be used to resum emissions to all orders. It 
leads to a Poisson distribution for visible emissions 
\begin{alignat}{3}
  \label{eq:poisson}
  \sigma_n \propto \frac{ \bar{n}^n}{n!} e^{-\bar{n}} \qquad \text{with} \quad
  \bar{n} \propto \frac{\alpha}{\pi} \; \text{log}
  \frac{E_{\text{hard}}}{E_{\text{soft}}} \; .
\end{alignat}
The numerator is just the exponentiation of $n$ emission
probabilities, while the $n!$ factor takes care of the bosonic phase
space. The exponential factor normalizes the distribution correctly.
This way we find the logarithmic dependence of $\bar{n}$, where
$E_\text{soft}$ is the minimum resolution for soft photons.  The
cross-section ratios for Poisson processes immediately follow as
\begin{align}
  \label{eq:ratios_poisson}
  R_{(n+1)/n} \equiv \frac{\sigma_{n+1} }{\sigma_n} = \frac{
    \bar{n} }{n+1} \; .
\end{align}
We observe this behavior in all QED processes in the soft limit.

\subsection{Staircase scaling}

In contrast to Eq.\eqref{eq:ratios_poisson} we find constant values
for QCD and $W/Z$ plus jets at hadron colliders. This behavior is
called staircase scaling and
follows~\cite{Englert:2011cg,Ellis:1985vn}
\begin{align}
  \label{eq:ratios_staircase}
  \phantom{haaaaaaaaaaaaaaaaallooooooooooooooooo}
  R_{(n+1)/n} \equiv \frac{\sigma_{n+1} }{\sigma_n} = R \; .
\end{align}
\vspace*{-8mm}\\
\begin{wrapfigure}{l}{0.45\textwidth}
  \vspace*{-8mm}
  \centering
  \includegraphics[width=0.45\textwidth]{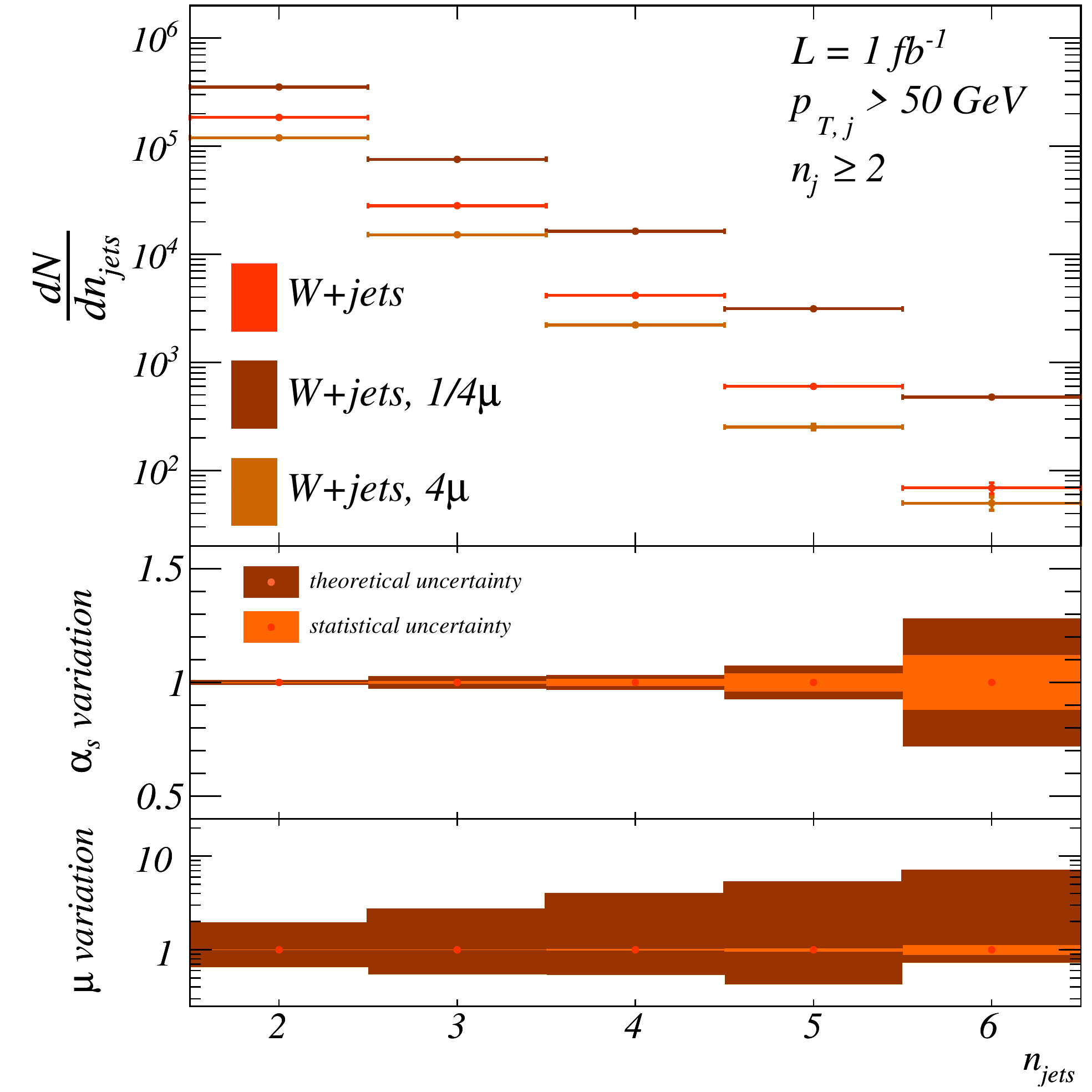}
  \caption{Theory uncertainties for $W$ plus jets production. Figure
    from Ref.\cite{Englert:2011cg}.}
  \label{fig:wjets}
  \vspace*{-5mm}
\end{wrapfigure}
Staircase scaling is a well established fact known since
UA1~\cite{Ellis:1985vn} and has been observed by ATLAS and
CMS~\cite{Aad:2010ab,cms}. Using {\sc Sherpa}~\cite{Gleisberg:2008ta}
we simulate exclusive $n_{\text{jets}}$ rates for $W/Z$ plus jets and
for QCD jets up to $n_{\text{jets}}=8$ and reproduce this pattern. A
major issue in the prediction of exclusive observables is the
estimation of theoretical uncertainties. We rely on two handles: the
value of the strong coupling $\alpha_S(m_Z)$ and a free overall scale
parameter connected to the factorization scale. The uncertainties we
estimate by varying $\alpha_s(m_Z)$ within its allowed values and by
multiplying the default scale by $1/4$ and $4$. In
Fig.~\ref{fig:wjets} we show the $n_{\text{jets}}$ distribution
including uncertainties for $W$ plus jets. While the variation of
$\alpha_s$ only gives a small error bar the impact of changing $\mu$
is very large. However, the actual staircase pattern is not altered.
Interpreting the large scale variation as an effect beyond the
expected accuracy we can treat it as a MC tuning parameter, which
happens to be close to unity for {\sc Sherpa}~\cite{Englert:2011cg}.

\section{Photon laboratory}

\begin{wrapfigure}{r}{0.35\textwidth}
  \centering
  \vspace*{-8mm}
  \includegraphics[width=0.35\textwidth]{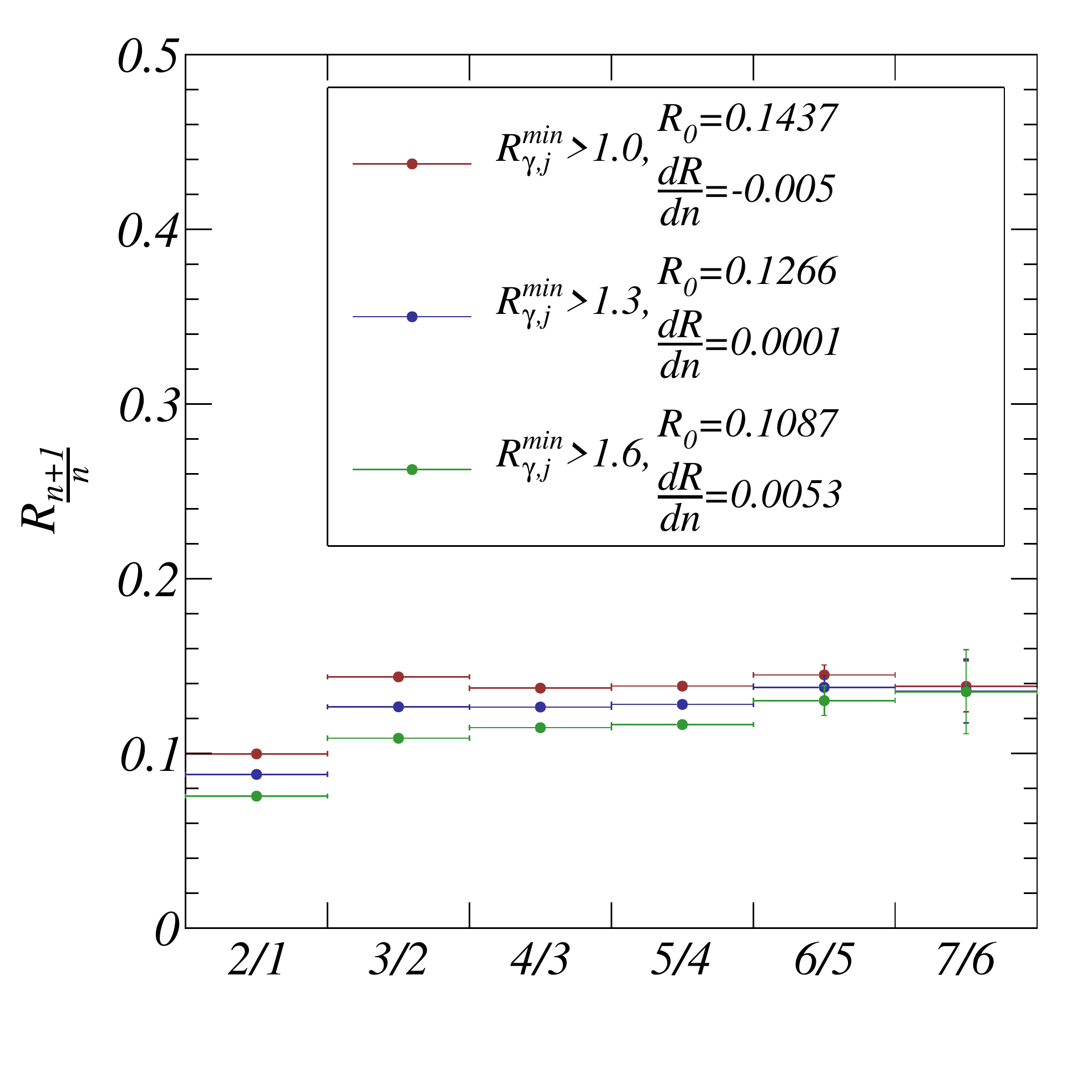}\\[-5mm]
  \includegraphics[width=0.35\textwidth]{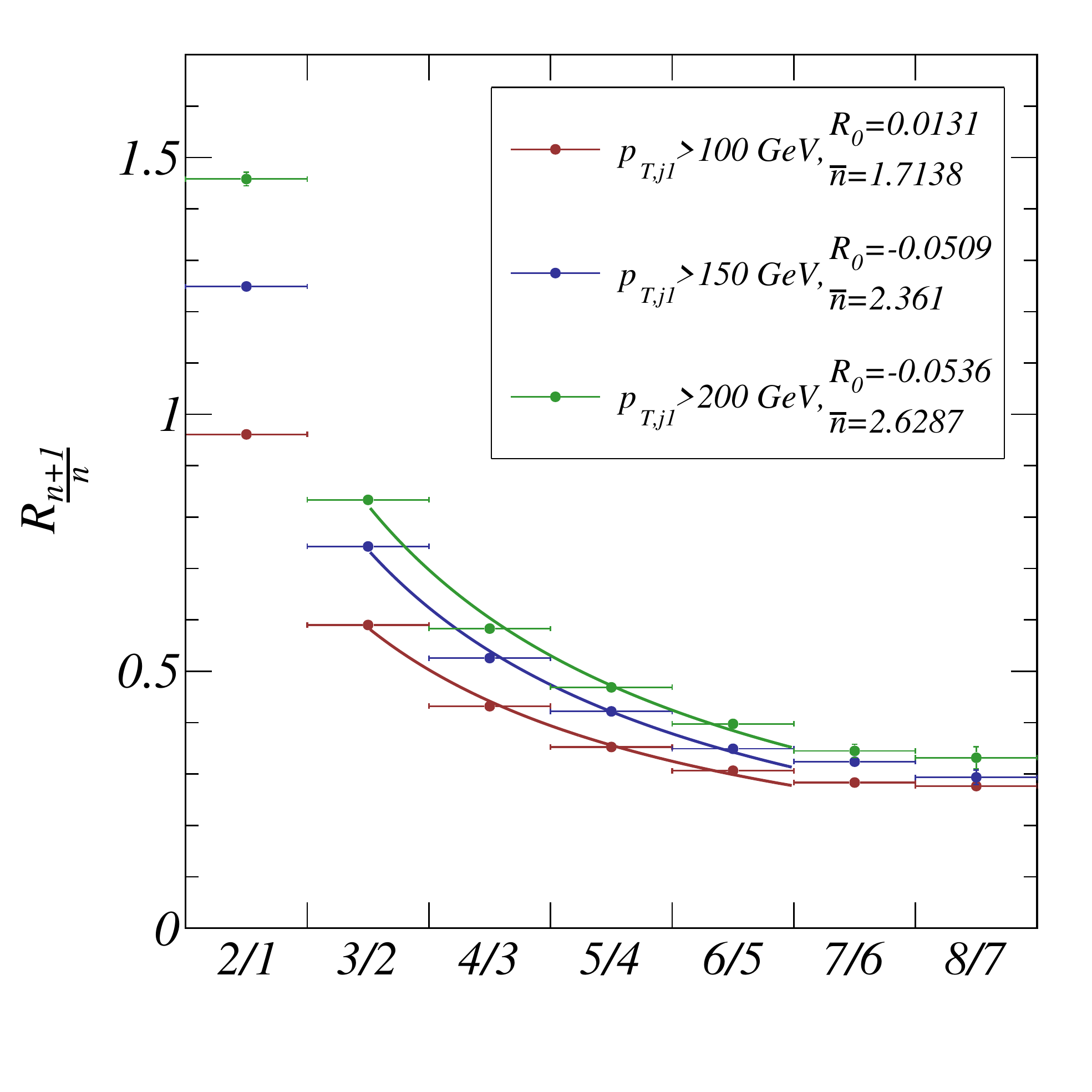}
  \vspace*{-9mm}
  \caption{Kinematic regimes showing staircase and Poisson
    scaling. Figure from Ref.~\cite{Englert:2011pq}.}
  \label{fig:photon}
  \vspace*{-25mm}
\end{wrapfigure}
The perfect place to study Poisson and staircase scaling in more
detail is photon plus jets~\cite{Englert:2011pq}. It has a high cross
section and is therefore accessible for early LHC data. At first
glance neither Poisson nor staircase scaling is observed in this
channel. In contrast to the $W/Z$ case the photon has no mass to
define a hard process.

Inspired by the staircase pattern in $W/Z$ plus jets we propose the
following cut scenario: count only jets and isolated photons above
$p_T^\text{min}$, then impose a wide separation cut between the photon
and all the counted jets either in terms of the invariant mass or
equivalently in terms of $R$. In Fig.~\ref{fig:photon} we observe
staircase scaling for values of $R>1.0$ and for invariant masses
around 90~GeV, given $p_T^{\text{min}}=50$~GeV.

To see Poisson scaling we induce a large logarithm as in
Eq.\eqref{eq:poisson} by asking for one jet with $p_T>100$ GeV and
lowering $p_T^{\text{min}}=20$~GeV. As we can see in
Fig.~\ref{fig:photon} the cross section ratios follow a Poisson
distribution. For high jet multiplicities the logarithm runs out of
steam and we return to staircase scaling, with a constant ratio $R$
determined by $p_T^{\text{min}}=20$~GeV. The quantitative description
of the staircase and Poisson scaling in the photon plus jets can be
directly linked to the $W/Z$ plus jets case~\cite{Englert:2011pq}.

\section{Applications}

\subsection{Higgs searches}

\begin{figure}[b]
  \centering
  \includegraphics[width=0.43\textwidth]{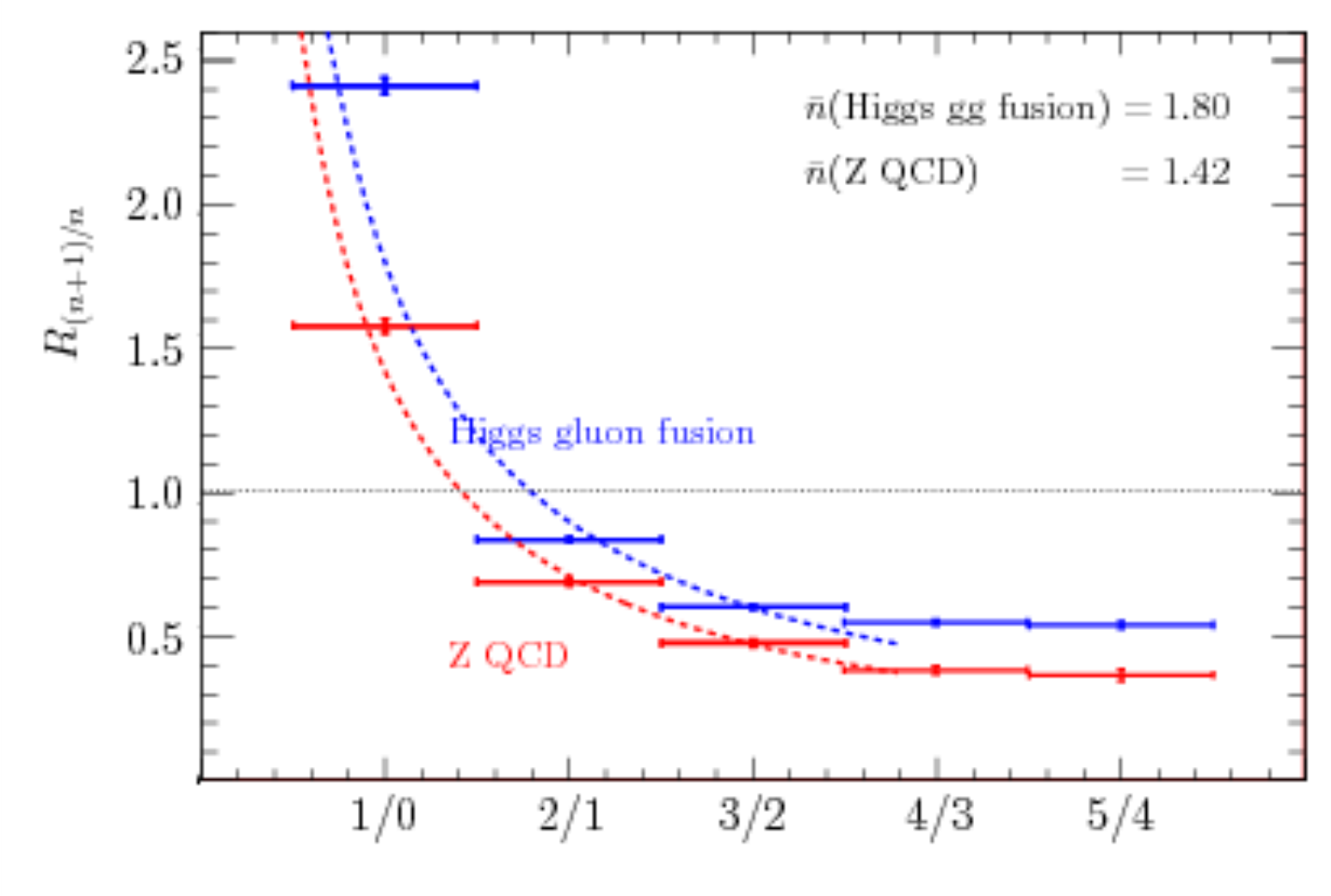}
  \hspace*{10mm}
  \includegraphics[width=0.43\textwidth]{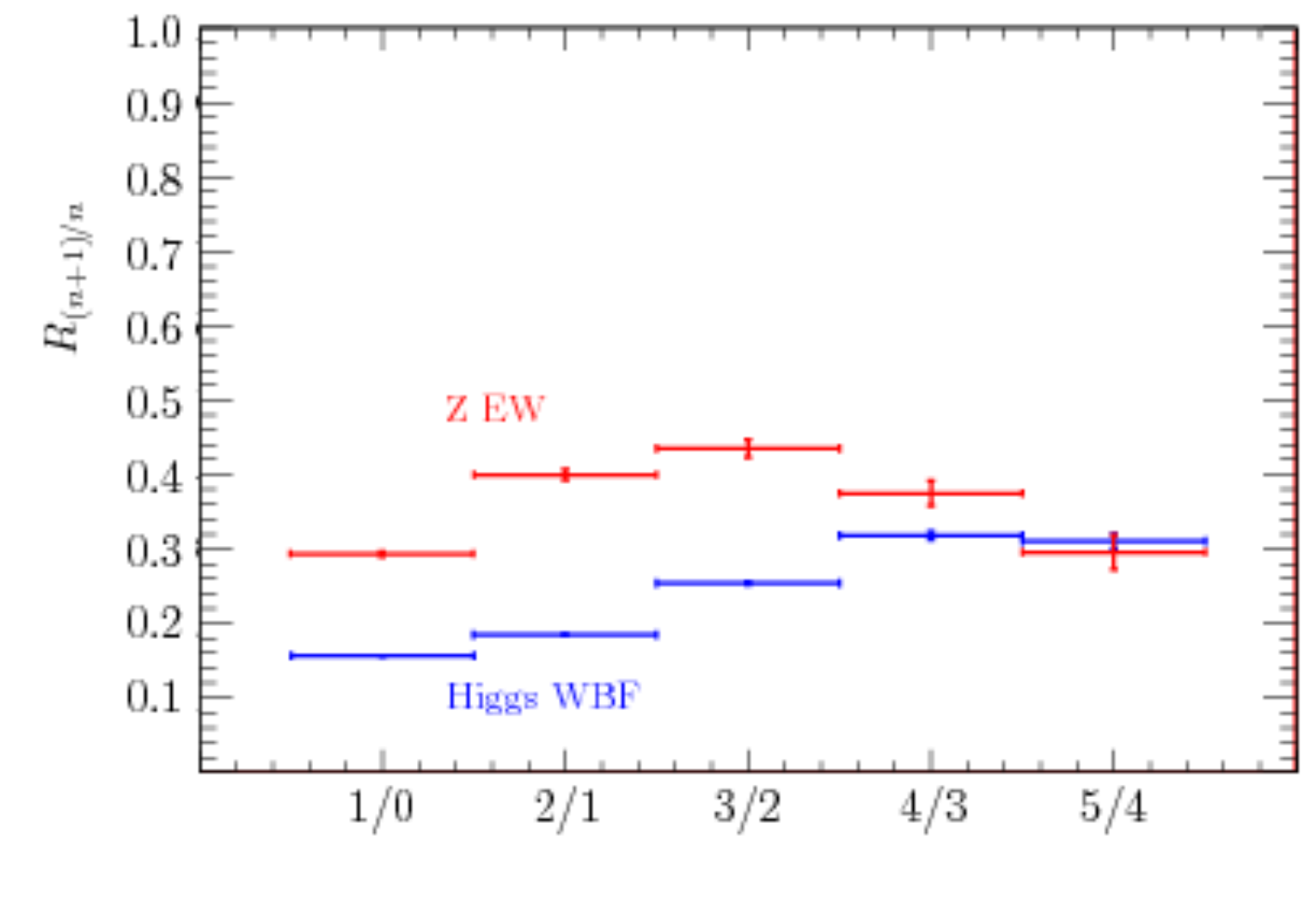}
  \vspace*{-5mm}
  \caption{Poisson backgrounds (left) and staircase signal (right) for
    Higgs production. Figure from Ref.~\cite{Gerwick:2011tm}.}
  \label{fig:higgs}
\end{figure}

In WBF Higgs searches we use a jet veto to suppress QCD
backgrounds. The prediction of the jet veto probability is
notorious~\cite{Gerwick:2011tm}. In Fig.~\ref{fig:higgs} we show how
the WBF cuts drive the backgrounds into the Poisson regime while the
signal stays approximately staircase. A simple fit to the
$n_\text{jets}$ distribution gives the veto survival probability.

\subsection{Inclusive searches and autofocus}

Searches for new physics focus on heavy colored states, for example
decaying to dark matter. Contrary to tuned cuts searches, which rely
on model spectra, we propose an inclusive ansatz, where we only count
jets and apply a missing energy cut~\cite{Englert:2011cg}. Information
about the heavy mass scale is encoded in the effective mass. Due to
its close connection to $n_\text{jets}$ this mass observable is well
controlled and can be used in our analysis. It yields complementary
information to the number of jets. While $n_\text{jets}$ is sensitive
to deviations mostly in the high multiplicity regime, the effective
mass also is sensitive for low multiplicities.

\begin{figure}[t]
  \centering
  \includegraphics[width=0.24\textwidth]{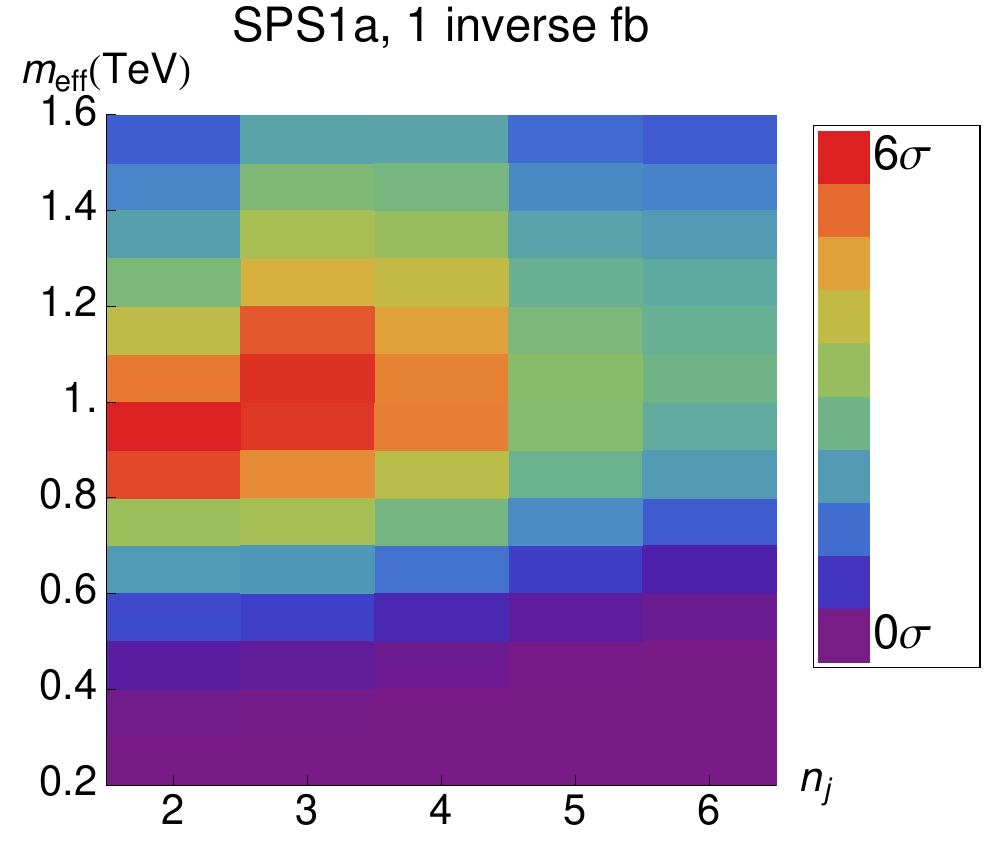}
  \includegraphics[width=0.24\textwidth]{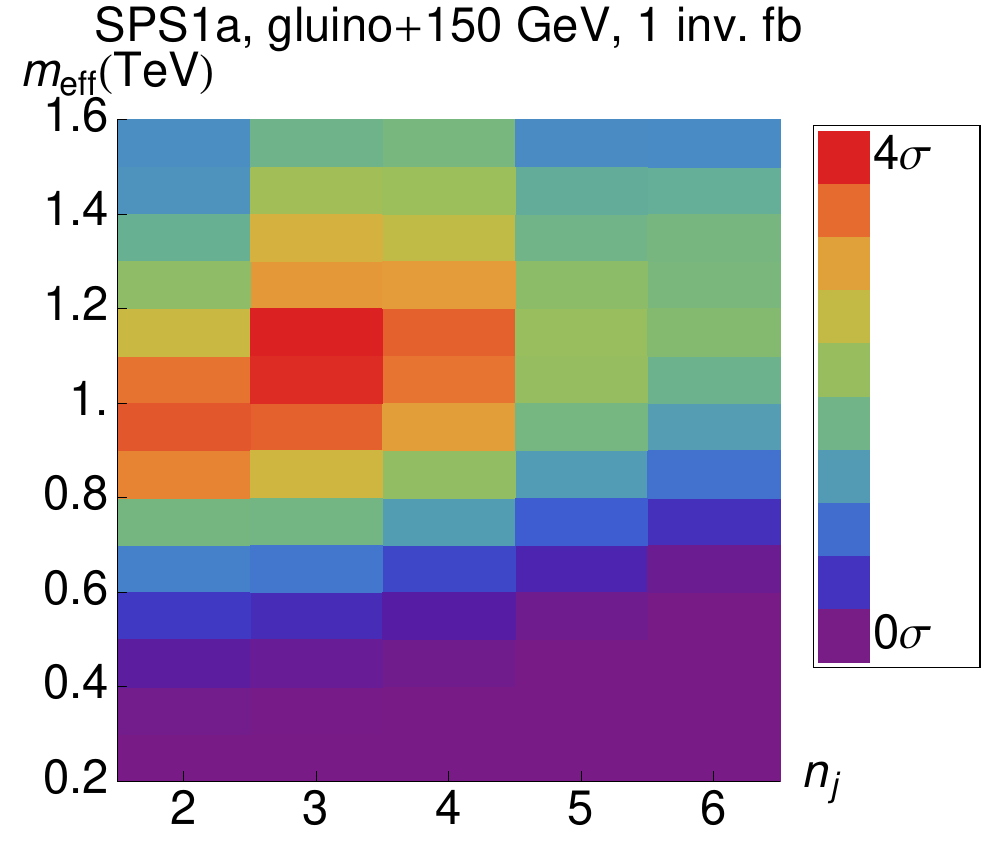}
  \includegraphics[width=0.24\textwidth]{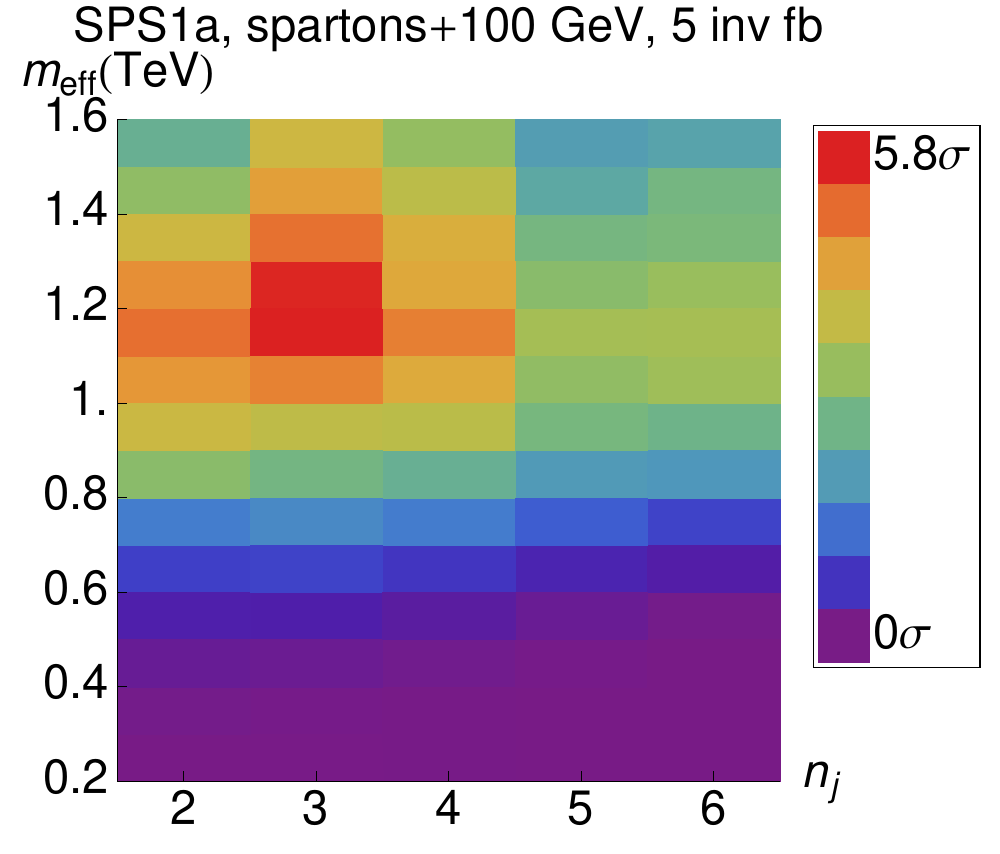}
  \includegraphics[width=0.24\textwidth]{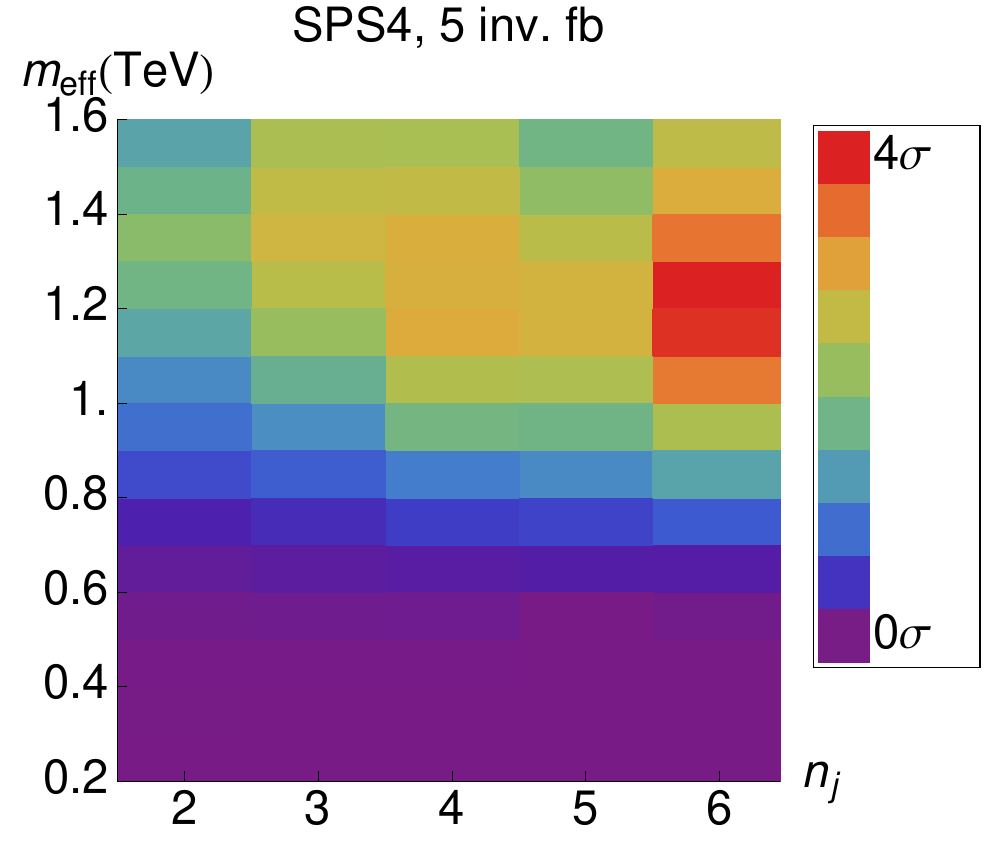}
  \caption{Two-dimensional likelihood for an SPS1a and SPS4 SUSY
    signal over backgrounds. Figure from Ref.~\cite{Englert:2011cg}.}
  \label{fig:auto}
\end{figure}

For a simple supersymmetric spectrum we use both observables to
perform a log-likelihood test of the SM and SUSY hypotheses. The
two-dimensional likelihoods for the different squark and gluino
channels we show in Fig.~\ref{fig:auto}. While the $m_\text{eff}$ axis
reflects the mass of the pair of heavy new states, the $n_\text{jets}$
axis is sensitive to the color charge of the squarks and gluinos.

\section{Acknowledgments}

P.S. acknowledges support by the International Max Planck Research
School for Precision Tests of Fundamental Symmetries.

% ****************************************************************************
% BIBLIOGRAPHY AREA
% ****************************************************************************

{\raggedright
\begin{footnotesize}
% IF YOU DO NOT USE BIBTEX, USE THE FOLLOWING SAMPLE SCHEME FOR THE REFERENCES
% ----------------------------------------------------------------------------
%\begin{thebibliography}{99}
%------- replace following references ;-)

%\bibitem{rainwater} D.~Rainwater {\it et al.} Phys.Rev. {\bf D60}
%  (1999) 11.

%\bibitem{review} D.~Morrissey {\it et al.} Phys.Rept. {\bf 515}
%  (2012) 1-113.

%\bibitem{sherpa} T.~Gleisberg {\it et~al.}  JHEP {\bf 0902} (2009) 007.

%\bibitem{ckkw} S.~Catani {\it et al.} JHEP {\bf 0111} (2001) 063.

%\bibitem{peskin} S.~Peskin ``Introduction to Quantumfield Theory''.

%\bibitem{laenen} E.~Laenen {\it et al} JHEP {\bf 1101} (2011) 141.

%\bibitem{autofocus} C.~Englert {\it et al.} Phys.Rev. {\bf D83} (2011)
%  095009.

%\bibitem{ellis} S.~Ellis {\it et al.} Phys.Lett. {\bf B154} (1985) 435.

%\bibitem{atlas} ATLAS Collaboration Phys.Lett. {\bf B698} (2011)
%  325-345.

%\bibitem{cms} CMS Collaboration CMS-PAS-EWK-10-012.

%\bibitem{photons} C.~Englert {\it et al.} JHEP {\bf 1202} (2012) 030.

%\bibitem{higgs} E.~Gerwick {\it et al.} Phys.Rev.Lett. {\bf 108} (2012)
%  032.

%\end{thebibliography}
% ----------------------------------------------------------------------------

% IF YOU USE BIBTEX,
% - DELETE THE TEXT BETWEEN THE TWO ABOVE DASHED LINES
% - UNCOMMENT THE NEXT TWO LINES AND REPLACE 'smith_joe.bib' WITH YOUR
%   FILE(S)

 \bibliographystyle{DISproc}
 \bibliography{schichtel_peter.bib}
\end{footnotesize}
}

% ****************************************************************************
% END OF BIBLIOGRAPHY AREA
% ****************************************************************************

\end{document}